\documentclass[preprint2]{aastex}
\usepackage{natbib}
\bibpunct{(}{)}{;}{a}{}{,}

\def\be{\begin{equation}}
\def\ee{\end{equation}}
\def\bea{\begin{eqnarray}}
\def\eea{\end{eqnarray}}

\def\l{\left}
\def\r{\right}
\def\ot{\frac{1}{2}}
\def\lra{\leftrightarrow}

\def\MeV{{~\rm MeV}}

\newcommand{\numu}{\nu_\mu}
\newcommand{\numutau}{\nu_{\mu,\tau}}
\newcommand{\nue}{\nu_{\rm e}}

\begin{document}

\title{{Electron Neutrino Pair Annihilation: A New Source \\
for Muon and Tau Neutrinos in Supernovae}}

\author{Robert Buras\altaffilmark{1,2}, Hans-Thomas
Janka\altaffilmark{1}, Mathias Th.~Keil\altaffilmark{2},\\ 
Georg G.~Raffelt\altaffilmark{2}, and 
Markus Rampp\altaffilmark{1}}

\affil{$^1$Max-Planck-Institut f\"ur Astrophysik, 
Karl-Schwarzschild-Str.~1, 85741 Garching, Germany}

\affil{$^2$Max-Planck-Institut f\"ur Physik 
(Werner-Heisenberg-Institut)\\ F\"ohringer Ring 6, 80805 M\"unchen,
Germany}

\begin{abstract}
We show that in a supernova core the annihilation process $\nu_{\rm
e}\bar\nu_{\rm e}\to\nu_{\mu,\tau}\bar\nu_{\mu,\tau}$ is always more
important than the traditional reaction ${\rm e}^+{\rm
e}^-\to\nu_{\mu,\tau}\bar\nu_{\mu,\tau}$ as a source for muon and tau
neutrino pairs.  We study the impact of the new process by means of a
Monte Carlo transport code with a static stellar background model and
by means of a self-consistent hydrodynamical simulation with Boltzmann
neutrino transport.  Nucleon bremsstrahlung ${\rm NN}\to {\rm
NN}\nu_{\mu,\tau}\bar\nu_{\mu,\tau}$ is also included as another
important source term.  Taking into account $\nu_{\rm e}\bar\nu_{\rm
e}\to\nu_{\mu,\tau}\bar\nu_{\mu,\tau}$ increases the $\nu_\mu$ and
$\nu_\tau$ luminosities by as much as 20\% while the spectra remain
almost unaffected.  In our hydrodynamical simulation the shock was
somewhat weakened.  Elastic $\nu_{\mu,\tau}\nue$ and
$\nu_{\mu,\tau}\bar\nue$ scattering is not negligible but less
important than $\nu_{\mu,\tau}{\rm e}^\pm$ scattering.  Its influence
on the $\nu_{\mu,\tau}$ fluxes and spectra is small after all other
processes have been included.
\end{abstract}


%%%%%%%%%%%%%%%%%%%%%%%%%%%%%%%%%%%%%%%%%%%%%%%%%%%%%%%%%%%%%%%%%%%%%%
%% Section I %%%%%%%%%%%%%%%%%%%%%%%%%%%%%%%%%%%%%%%%%%%%%%%%%%%%%%%%%
%%%%%%%%%%%%%%%%%%%%%%%%%%%%%%%%%%%%%%%%%%%%%%%%%%%%%%%%%%%%%%%%%%%%%%

\section{\uppercase{Introduction}}

\label{sec:Introduction}

The treatment of $\nu_\mu$ and $\nu_\tau$ transport in numerical
supernova (SN) simulations has been somewhat schematic in the
past. However, with the advent of numerical Boltzmann solvers for the
neutrino transport (Mezzacappa \& Bruenn 1993, Mezzacappa \& Messer
1999; Yamada, Janka, \& Suzuki 1999; Burrows et al.\ 2000; Rampp \&
Janka 2002) and their application to the post-bounce phase of
stellar core-collapse models (Rampp \& Janka 2000; Mezzacappa et~al.\
2001; Liebend\"orfer et al.\ 2001) a new level of accuracy has been
achieved. Evidently it is desirable that in consistent
state-of-the-art simulations the uncertainties are not dominated by
overly crude approximations of the microphysics which governs the
neutrino interactions. For example, it has been recognized that
nuclear many-body correlations (Burrows \& Sawyer 1998; Reddy et~al.\
1999) or weak-magnetism effects in neutrino-nucleon interactions
(Vogel \& Beacom 1999; Horowitz \& Li 2000; Horowitz 2002) should be
included.

Previous simulations used iso-energetic scattering on nucleons
$\nu_\mu{\rm N}\to{\rm N}\nu_\mu$ as the main opacity source for
$\nu_\mu$ transport, elastic scattering on electrons and positrons
$\nu_\mu {\rm e}^\pm\to {\rm e}^\pm\nu_\mu$ as the main
energy-exchange reaction, and ${\rm e}^+{\rm
e}^-\to\nu_\mu\bar\nu_\mu$ as the only source term for $\nu_{\mu}$
production.  (Here and in what follows we use $\nu_\mu$ symbolically
for either $\nu_\mu$ or $\nu_\tau$.)  However, it is now generally
accepted that nucleon bremsstrahlung ${\rm NN}\to {\rm
NN}\nu_{\mu}\bar\nu_{\mu}$ is important or even dominant as a neutrino
source reaction (Suzuki 1991,1993; Hannestad \& Raffelt 1998;
Thompson, Burrows, \& Horvath 2000), and that nucleon recoils have a
significant impact on the emerging $\nu_\mu$ flux spectrum (Janka
et~al.\ 1996; Raffelt 2001; Keil, Raffelt, \& Janka 2002).

\begin{figure}[ht]
\columnwidth=4.5cm
\plotone{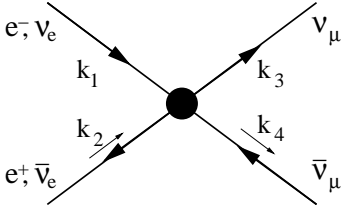}
\caption{\label{fig:feyn} Feynman graph for the annihilation
processes producing $\nu_\mu\bar\nu_\mu$ pairs.}
\end{figure}

In this paper we show that in addition $\nu_{\rm e}\bar\nu_{\rm
e}\to\nu_{\mu}\bar\nu_{\mu}$ and its inverse reaction should be
included because this process is always far more important than ${\rm
e}^+{\rm e}^-\to\nu_{\mu}\bar\nu_{\mu}$ as a neutrino source
(Fig.~\ref{fig:feyn}).
Conversely, $\nu_{\mu}\nu_{\rm e}$ and $\nu_{\mu}\bar\nu_{\rm e}$
scattering turns out to be less important than $\nu_{\mu}{\rm e}^\pm$
scattering and thus not crucial for the $\nu_\mu$ flux and spectra
formation.  Bond (1978) has previously discussed these processes in
the context of neutrino transport in supernovae, but no assessment of
their relative importance has been performed.

It is straightforward to see that $\nu_{\rm e}\bar\nu_{\rm
e}\to\nu_{\mu}\bar\nu_{\mu}$ must be important in a SN core. The
region of interest is well inside the $\nu_{\mu}$ transport sphere,
i.e.\ interior to the radius where $\nu_\mu{\rm N}\to{\rm N}\nu_\mu$
scattering freezes out. This region is always deeper in the star than
the freeze-out radius of the charged-current reactions $\nu_{\rm
e}{\rm n}\lra {\rm p}{\rm e}^-$ and $\bar\nu_{\rm e}{\rm p}\lra {\rm
n}{\rm e}^+$, implying that in the $\nu_\mu$ trapping region $\nu_{\rm
e}$ and $\bar\nu_{\rm e}$ are essentially in local thermodynamic
equilibrium (LTE).  Therefore, in this region the process $\nu_{\rm
e}\bar\nu_{\rm e}\leftrightarrow \nu_{\mu}\bar\nu_{\mu}$ will serve to
create or destroy $\nu_{\mu}\bar\nu_{\mu}$ pairs in the same way as
the traditional ${\rm e}^+{\rm
e}^-\leftrightarrow\nu_{\mu}\bar\nu_{\mu}$ process.  With vanishing
chemical potentials of ${\rm e}^-$ and $\nu_{\rm e}$, the rate of the
new process turns out to be about twice that of the traditional one so
that the combined source strength would be three times the traditional
one.

\begin{figure}[b]
\columnwidth=7.5cm
\plotone{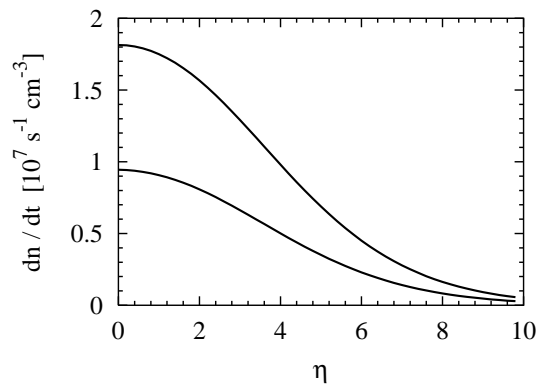}
\caption{\label{fig:easyexpl} Pair production rates by the process
$\nu_{\rm e}\bar\nu_{\rm e}\to\nu_{\mu}\bar\nu_{\mu}$ as a function of
$\eta_{\nu_{\rm e}}$ (upper line) and ${\rm e}^+{\rm
e}^-\to\nu_{\mu}\bar\nu_{\mu}$ as a function of $\eta_{\rm e}$ (lower
line).  We used $T=12~\MeV$ and $\eta_{\nu_\mu}=0$.}
\end{figure}

In the presence of non-vanishing chemical potentials the enhancement
is even larger. The pair production rate due to ${\rm e}^+{\rm
e}^-\to\nu_{\mu}\bar\nu_{\mu}$ is shown in Fig.~\ref{fig:easyexpl} as
a function of the degeneracy parameter $\eta_{\rm e}=\mu_{\rm e}/T$
for the electrons, and likewise, the rate for $\nu_{\rm e}\bar\nu_{\rm
e}\to\nu_{\mu}\bar\nu_{\mu}$ as a function of $\eta_{\nu_{\rm
e}}$. The production rate is a decreasing function of $\eta$. Since in
the relevant regions of the SN core $\eta_{\nu_{\rm e}}<\eta_{{\rm
e}}$, the traditional process of ${\rm e}^+{\rm e}^-$ annihilation is
reduced more strongly than the new $\nue\bar\nue$ rate. The latter
therefore dominates even more.

In order to estimate the impact of the new source reaction on the
neutrino fluxes and spectra we conduct two separate numerical
investigations. First, we perform a Monte Carlo transport simulation
on a static stellar background model. While this approach has the
disadvantage of not following the evolution of the neutron star
atmosphere self-consistently, it enables us to disentangle the
individual effects of different neutrino processes on the transport
and spectra formation in a systematic study.  Next, we perform two
full-scale numerical post-bounce simulations where once the new
process is included and once it is left out.  This allows us to verify
that the effects established by the Monte Carlo results are generic
and also show up in self-consistent radiation-hydrodynamical
models. The use of these two different approaches and independent
codes also helps making sure that our results do not depend on details
of the technical implementation or the particular numerical
resolution.

We begin in Sec.~\ref{sec:Comparison} by comparing the ${\rm e}^+{\rm
e}^-$ and $\nu_{\rm e}\bar\nu_{\rm e}$ reactions. In
Sec.~\ref{sec:MonteCarlo} we discuss the results of a Monte Carlo
study of neutrino transport while in Sec.~\ref{sec:Boltzmann} we
describe the self-consistent hydrodynamical simulations coupled with a
Boltzmann transport solver.  We summarize our findings in
Sec.~\ref{sec:Conclusions}.

%%%%%%%%%%%%%%%%%%%%%%%%%%%%%%%%%%%%%%%%%%%%%%%%%%%%%%%%%%%%%%%%%%%%%%
%% Section II %%%%%%%%%%%%%%%%%%%%%%%%%%%%%%%%%%%%%%%%%%%%%%%%%%%%%%%%
%%%%%%%%%%%%%%%%%%%%%%%%%%%%%%%%%%%%%%%%%%%%%%%%%%%%%%%%%%%%%%%%%%%%%%

\begin{deluxetable}{lll}
\tablecaption{\label{tab:CVA}
Weak interaction constants.}
\tablewidth{0pt}
\tablehead{Process:
&${\rm e}^+ {\rm e}^-\lra\nu_\mu\bar\nu_\mu$
&$\nu_{\rm e} \bar\nu_{\rm e}\lra\nu_\mu\bar\nu_\mu$}
\startdata
$C_{\rm V}$&  $-\ot+2\sin^2\theta_w$ & $+\ot$\\
\noalign{\medskip}
$C_{\rm A}$&  $-\ot$ & $+\ot$\\
\enddata
\end{deluxetable}

\section{\uppercase{Electron vs.\ Neutrino Pair Annihilation}}

\label{sec:Comparison}

We begin by comparing the two pair annihilation processes
\bea\label{eq:nunu}
\nu_{\rm e}+\bar\nu_{\rm e}&\to&\nu_\mu+\bar\nu_\mu\,,\\
\label{eq:eenunu}
{\rm e}^++{\rm e}^-&\to&\nu_\mu+\bar\nu_\mu\,.
\eea
In the relativistic limit where the electron mass can be neglected,
their squared and spin-summed matrix elements are of the form
\bea 
\sum_{\rm spins}\l|{\cal M}\r|^2 &=&8~G_F^2 \l[\r.(C_{\rm V}+C_{\rm A})^2u^2 
\nonumber\\
&&\kern2em{}+(C_{\rm V}-C_{\rm A})^2t^2\l.\r],
\label{eq:matrixel} 
\eea
with the Mandelstam variables $t=-2k_1\cdot k_3$ and $u=-2k_1\cdot
k_4$. The momenta are assigned to the particles as indicated in
Fig.~\ref{fig:feyn}.  The matrix elements for the two processes differ
only in the weak coupling constants $C_{\rm V,A}$ shown in
Table~\ref{tab:CVA}.

The pair production rate is obtained by appropriate phase-space
integrations, including particle distributions and blocking factors
(Yueh \& Buchler 1976; Hanne\-stad \& Madsen 1995). We use
$\mu_{\nu_\mu}=0$ even though there could be a small $\nu_\mu$
chemical potential due to a non-vanishing concentration of muons in
the core and due to different transport properties of $\nu_\mu$ and
$\bar\nu_\mu$.  Assuming that ${\rm e}^\pm$, $\nu_{\rm e}$, and
$\bar\nu_{\rm e}$ are all in LTE, the rates of pair creation vs.\
degeneracy parameters $\eta_{\rm e}$ and $\eta_{\nu_{\rm e}}$ are
shown in Fig.~\ref{fig:easyexpl}.  In the dense regions of a SN core
below the neutrino spheres the phase space distribution of electron
neutrinos is a Fermi-Dirac function with degeneracy parameter
$\eta_{\nue}=\eta_{\rm e}+\eta_{\rm p} - \eta_{\rm n} < \eta_{\rm e}$
so that the new process is always more important than ${\rm e}^+{\rm
e}^-$ annihilation.

\begin{figure}[b]
\columnwidth=7.5cm
\plotone{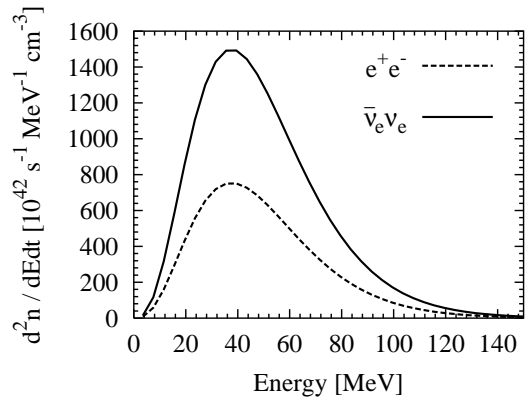}
\caption{\label{fig:energy-rates-eta0} Differential $\nu_\mu$ and
$\bar\nu_\mu$ production rates ${\rm d}^2n/{\rm d}\epsilon{\rm d}t$
vs.\ neutrino energy for $\eta_{\rm e}=\eta_{\nu_{\rm
e}}=0$ and $T=12\MeV$.}
\end{figure}

An interesting difference between the two processes arises in the
differential production rates, i.e.\ the $\nu_\mu$ and $\bar\nu_\mu$
production rates as functions of neutrino energy $\epsilon$.  As a
first case we take $\eta=0$ for both e and $\nu_{\rm e}$ and show the
differential production rate ${\rm d}^2 n/{\rm d}\epsilon{\rm d}t$ in
Fig.~\ref{fig:energy-rates-eta0}. For both processes the differential
rate is the same for $\nu_\mu$ and $\bar\nu_\mu$, i.e.\ they are
produced with the same spectra.

\begin{figure}[ht]
\columnwidth=7.5cm
\plotone{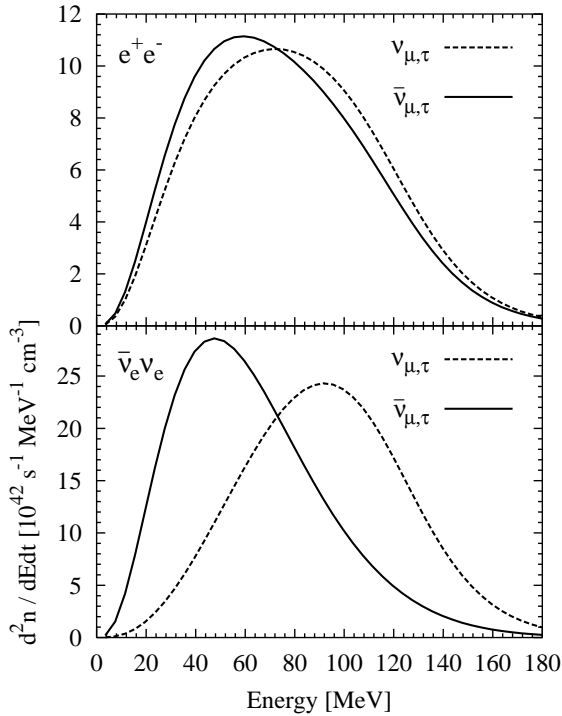}
\caption{\label{fig:energy-rates-eta10} Differential $\nu_\mu$ and
$\bar\nu_\mu$ production rates ${\rm d}^2n/{\rm d}\epsilon{\rm d}t$ 
for $\eta_{\rm e}=\eta_{\nu_{\rm e}}=10$
and $T=12\MeV$.  Upper panel for ${\rm e}^+{\rm
e}^-\to\nu_\mu\bar\nu_\mu$, lower panel for $\nu_{\rm
e}\bar\nu_{\rm e}\to\nu_\mu\bar\nu_\mu$.}
\end{figure}

However, in general the ``parent particles'' will have a significant
chemical potential. Taking $\eta_{\rm e}=\eta_{\nu_{\rm e}}=10$ we
show the differential production rates in
Fig.~\ref{fig:energy-rates-eta10}. In case of the ${\rm e}^+{\rm e}^-$
process (upper panel) the differential rates are similar for $\nu_\mu$
and $\bar\nu_\mu$. This is understood by the fact that the values of
$(C_{\rm V}+C_{\rm A})^2\simeq 0.54^2$ and $(C_{\rm V}-C_{\rm
A})^2\simeq 0.46^2$ are quite similar so that the $u^2$ and $t^2$
terms in Eq.~(\ref{eq:matrixel}) are almost equally
important. Therefore, interchanging $\nu_\mu$ and $\bar\nu_\mu$,
corresponding to an exchange of $u$ and $t$, has no big effect.

This is not true for the neutrino reaction, where $(C_{\rm V}+C_{\rm
A})^2=1$ and $(C_{\rm V}-C_{\rm A})^2=0$ and thus only $u^2$
contributes. Replacing $u$ by $t$ now changes the kinematics of the
process, which in turn modifies the rate if the distribution of $\nue$
differs from that of $\bar\nue$, i.e.\ if $\eta_{\nue}\neq 0$. The
$\nu_\mu$ and $\bar\nu_\mu$ spectra shown in the bottom panel of
Fig.~\ref{fig:energy-rates-eta10} are indeed very different from each
other, although the total production rates of $\nu_\mu$ and
$\bar\nu_\mu$ are, of course, equal.

One can easily understand why $\nu_\mu$ on average have larger
energies than $\bar\nu_\mu$.  We first look at $\nu_{\rm
e}\bar\nu_{\rm e}\to\nu_\mu\bar\nu_\mu$ in the center of momentum (CM)
frame.  The differential cross section is \be \frac{{\rm
d}\sigma}{{\rm d} \cos \theta} = \frac{G_F^2}{4 \pi}\,\epsilon^2
(1+\cos\theta)^2 \,, \ee where $\theta$ is the angle between the
ingoing $\nue$ and the outgoing $\nu_\mu$, or equivalently, between
the ingoing $\bar\nue$ and the outgoing $\bar\nu_\mu$. Put another
way, forward scattering is favored and backward scattering
forbidden. This is due to angular momentum conservation. The ingoing
$\nu_{\rm e}$ and $\bar\nu_{\rm e}$ have opposite helicities and, in
the CM frame, opposite momenta, so that their combined spins add up to
1. The same is true for the outgoing particles so that backward
scattering would violate angular momentum conservation.  In the rest
frame of the medium the ingoing $\nue$ tends to have energies of the
order of its Fermi energy, while the ingoing $\bar\nue$ tends to have
energies of order $T$. Because forward scattering is favored, the
outgoing $\nu_\mu$ tends to inherit the larger energy of the ingoing
$\nu_{\rm e}$.

The differences of the source spectra, however, do not translate into
significant spectral differences of the $\nu_\mu$ and $\bar\nu_\mu$
fluxes emitted from the SN core. While pair annihilations and nucleon
bremsstrahlung are responsible for producing or absorbing neutrino
pairs and thus their equilibration with the stellar medium below the
``neutrino-energy sphere,'' other processes, notably $\nu_\mu {\rm
e}^\pm$ scattering and nucleon recoils, are more efficient for the
exchange of energy between neutrinos and the medium between the
equilibration and transport spheres.  In our numerical runs we will
find in fact that adding the new process to a SN simulation primarily
modifies the flux with only minor modifications of the spectrum.

\begin{figure}[b]
\columnwidth=7.5cm
\plotone{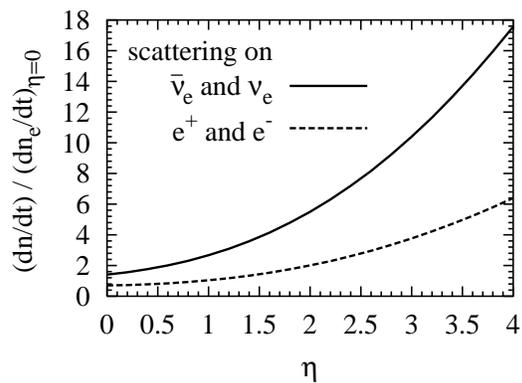}
\caption{\label{fig:easyexpl_scat} Thermally averaged scattering rate
for $\nu_\mu$ on ${\rm e}^\pm$ as a function of $\eta_{\rm e}$ (lower
line) and for $\nu_\mu$ on $\nue$ and $\bar\nue$ (upper line) as a
function of $\eta_{\nue}$. The rates are normalized to the scattering
rate on ${\rm e}^\pm$ at $\eta_{\rm e}=0$.  We used $T=12~\MeV$ and
$\eta_{\nu_\mu}=0$.}
\end{figure} 

If $\nue\bar\nue\to\numu\bar\numu$ is important relative to ${\rm
e}^+{\rm e}^-\to\numu\bar\numu$ one may wonder if processes of the
form 
\bea \numu + \nue&\to&\numu+\nue\,,\\
\numu+\bar\nue&\to&\numu+\bar\nue \,, 
\eea 
could be of comparable importance as $\nu_\mu{\rm e}^\pm$ scattering.
Figure~\ref{fig:easyexpl_scat} shows the rates for $\numu$ scattering
on $\nue$ and $\bar\nue$, and those for scattering on ${\rm e}^+$ and
${\rm e}^-$ as functions of $\eta_{\nue}$ and $\eta_{\rm e}$,
respectively.  The rates are normalized to the $\numu{\rm e}^\pm$ rate
at $\eta_{\rm e}=0$. In contrast to the annihilation rates, the
scattering rates rise monotonically with $\eta$.  Therefore, even
though neutrino-neutrino scattering would dominate if all chemical
potentials were zero, for realistic situations with
$\eta_{\nue}<\eta_{\rm e}$ we expect that scattering on ${\rm
e}^\pm$ has 1--2 times the rate of scattering on $\nue$ and
$\bar\nue$. Therefore, neutrino-neutrino scattering is expected to be
a relatively minor correction.  In our Monte Carlo studies we will
indeed find that this process has only a small effect on the neutrino
spectra and fluxes.

%%%%%%%%%%%%%%%%%%%%%%%%%%%%%%%%%%%%%%%%%%%%%%%%%%%%%%%%%%%%%%%%%%%%%%
%% Section III %%%%%%%%%%%%%%%%%%%%%%%%%%%%%%%%%%%%%%%%%%%%%%%%%%%%%%%
%%%%%%%%%%%%%%%%%%%%%%%%%%%%%%%%%%%%%%%%%%%%%%%%%%%%%%%%%%%%%%%%%%%%%%

\section{\uppercase{Monte Carlo Study}}

\label{sec:MonteCarlo}

To study the impact of the new annihilation process on the $\nu_\mu$
fluxes and spectra we first use a Monte Carlo method for neutrino
transport on a static background stellar model. For this purpose we
have adapted the Monte Carlo code of Janka \& Hillebrandt (1989a,b) to
include additional processes and nucleon recoil. We use this code for
an extensive parameter study of $\nu_\mu$ spectra formation that will
be documented elsewhere (Keil et.~al.\ 2002).

\begin{figure}[ht]
\columnwidth=7.5cm
\plotone{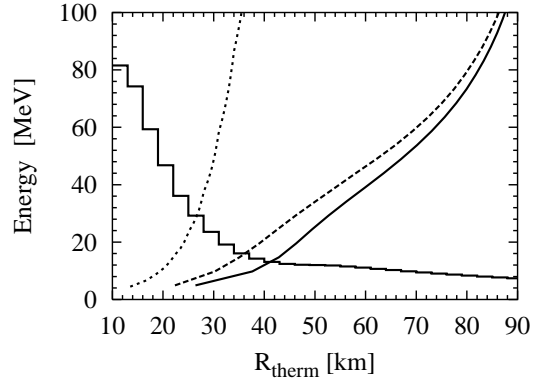}
\caption{\label{fig:rtherm}Thermalization depth $R_{\rm therm}$
(horizontal axis) for different processes as a function of 
neutrino energy $\epsilon$
(vertical axis) in our background model.  The energy-exchanging
processes are $\nu_{\rm e}{\rm n}\to {\rm p}{\rm e}^-$ (solid line),
$\bar\nu_{\rm e}{\rm p}\to {\rm n}{\rm e}^+$ (dashed line), and
$\nu_\mu\bar\nu_\mu$ annihilation to $\nu_{\rm e}\bar\nu_{\rm e}$
pairs (dotted line). The steps represent the temperature profile
of the stellar model in terms of 
$\langle\epsilon\rangle\simeq 3.15\,T$.}
\end{figure}

As a stellar background we employ a model originally provided to us by
B.~Messer for an earlier study of neutrino spectra formation (Raffelt
2001). Using this model again for our present study facilitates a
comparison with this previous work. The model is based on a full-scale
Newtonian collapse simulation of the Woosley \& Weaver $15\,M_\odot$
progenitor model labeled s15s7b.  We use a snapshot at 324~ms after
bounce when the shock wave is at a radius of about 120~km, i.e.\ the
SN core still accretes matter.  In Fig.~\ref{fig:rtherm} the
temperature profile is represented by the steps in terms of
$\langle\epsilon\rangle\simeq3.15\,T$ for each radial zone, i.e.\ by
the average energy of nondegenerate neutrinos ($\eta_{\numu} = 0$) in
LTE with the stellar medium.

In Fig.~\ref{fig:rtherm} we also show the thermalization depth $R_{\rm
therm}$ for several processes as a function of neutrino energy
$\epsilon$. The formal definition of $R_{\rm therm}$ in terms of
an effective mean free path for energy exchange is given by the
condition (cf.\ Shapiro \& Teukolsky 1983; Suzuki 1989; 
Raffelt 2001)
\begin{equation}
\tau_{\rm eff}(\epsilon)\,=\,
\int\limits_{R_{\rm therm}}^\infty\!\!\!\!{\rm d}r\,
\sqrt{{1\over \lambda_i} \sum_{j}{1\over \lambda_j}}\,=\,  {2\over 3}
\end{equation}
for the effective optical depth $\tau_{\rm eff}$ of a particular
equilibrating process with mean free path $\lambda_i(\epsilon)$ among
the opacity producing reactions having mean free paths
$\lambda_j(\epsilon)$. The solid line in Fig.~\ref{fig:rtherm} is for
$\nu_{\rm e}{\rm n}\to {\rm p}{\rm e}^-$, i.e.\ it represents the
energy-dependent $\nu_{\rm e}$ sphere. The dashed line is the
analogous $\bar\nu_{\rm e}$ sphere due to $\bar\nu_{\rm e}{\rm p}\to
{\rm n}{\rm e}^+$. Finally, the dotted line shows where $\nu_\mu$ or
$\bar\nu_\mu$ of given energy last participate in the
$\nu_\mu\bar\nu_\mu\to \nu_{\rm e}\bar\nu_{\rm e}$ process, assuming
that the annihilation partners are distributed according to LTE.  Put
another way, the dotted line is the energy-dependent freeze-out sphere
for our new process.  It is always at much smaller radii than the
$\nu_{\rm e}$ and $\bar\nu_{\rm e}$ spheres. Therefore, in those
regions where $\nu_\mu\bar\nu_\mu\leftrightarrow \nu_{\rm
e}\bar\nu_{\rm e}$ is effective we may assume LTE for $\nu_{\rm e}$
and $\bar\nu_{\rm e}$.

We therefore implemented the new process in our Monte Carlo code by
using the same subroutine as for the ${\rm e}^+{\rm e}^-$ pair
process, except for inserting the appropriate weak interaction
constants $C_{\rm V,A}$ and replacing $\eta_{\rm e}$ by
$\eta_{\nu_{\rm e}}$.

\begin{deluxetable}{lllllccccrc}
\tablecaption{\label{tab:NumericalResultsMesser}
Spectral characteristics of neutrino fluxes from 
Monte Carlo transport.}
\tablewidth{0pt}
\tablehead{\multicolumn{5}{l}{Energy exchange}
&$\langle\epsilon\rangle_{\rm flux}$&
$\langle\epsilon^2\rangle_{\rm flux}$&$p_{\rm flux}$
&$T$&$\eta$&$L_\nu$\\
\multicolumn{5}{l}{}
&[MeV]&[MeV$^2$]&&[MeV]&&$[10^{51}~\rm \frac{erg}{s}]$}
\startdata
\multicolumn{11}{l}{$\nu_\mu$ transport}\\
\multicolumn{5}{l}{original run}& 17.5& 388.& 0.97&  5.2&  1.1& 14.4\\
--&--&s &p &--& 16.6& 362.& 1.01&  5.3&$-$0.3& 15.8\\
--&--&s &p &n & 16.9& 369.& 0.99&  5.3&  0.4& 20.2\\
b &r &s &p &--& 14.2& 255.& 0.98&  4.2&  1.1& 14.8\\
b &r &s &p &n & 14.4& 264.& 0.97&  4.3&  1.2& 17.6\\
b &r &s &--&n & 14.4& 263.& 0.97&  4.3&  1.2& 17.0\\
b &r &sn&p &n & 14.3& 260.& 0.97&  4.3&  1.2& 17.9\\
\multicolumn{11}{l}{~}\\
\multicolumn{11}{l}{$\bar\nu_\mu$ transport}\\
--&--&s &p &n & 16.9& 368.& 0.99&  5.2&  0.6& 20.6\\
b &r &s &p &n & 14.4& 263.& 0.97&  4.2&  1.3& 17.8\\
b &r &s &--&n & 14.4& 262.& 0.98&  4.3&  1.1& 16.8\\
\enddata
\tablecomments{For energy exchange, ``b'' refers to bremsstrahlung,
``r'' to recoil, ``s'' to scattering on electrons and positrons, 
``p'' to ${\rm e}^+ {\rm e}^-$ pair annihilation, ``n'' to 
$\nu_{\rm e}\bar\nu_{\rm e}$  pair annihilation, and ``sn'' to
scattering on both, ${\rm e}^\pm$ and $\nu_{\rm e}$, $\bar\nu_{\rm
e}$.} 
\end{deluxetable}

We summarize the characteristics of the emerging neutrino flux for our
runs in Table~\ref{tab:NumericalResultsMesser}. As introduced by Janka
\& Hillebrandt (1989a) we characterize the neutrino spectrum by its
first two energy moments $\langle\epsilon\rangle_{\rm flux}$ and
$\langle\epsilon^2\rangle_{\rm flux}$ and define the ``pinching
parameter'' by following Raffelt (2001) as
\begin{equation}
p_{\rm flux}\equiv \frac{1}{a}\,
\frac{\langle\epsilon^2\rangle_{\rm flux}}
{\langle\epsilon\rangle_{\rm flux}^2}\,,
\end{equation}
with the energy moments 
\begin{equation}
\langle\epsilon^n\rangle_{\rm flux}\,=\,
{ \int_0^\infty \!{\rm d}\epsilon\int_{-1}^{+1}\!{\rm d}\mu\,
f_{\nu}(\epsilon,\mu)\epsilon^{n+2}\mu   \over
\int_0^\infty \!{\rm d}\epsilon\int_{-1}^{+1}\!{\rm d}\mu\,   
f_{\nu}(\epsilon,\mu)\epsilon^2\mu }
\end{equation}
of the emergent flux spectrum. Here $f_{\nu}(\epsilon,\mu)$ is
the neutrino distribution function in energy-angle space with
$\mu$ being the cosine of the angle of neutrino propagation relative
to the radial direction. 
The constant $a$ for a Fermi-Dirac distribution at zero chemical
potential is
\begin{equation}
a\equiv\frac{\langle\epsilon^2\rangle}{\langle\epsilon\rangle^2}
=\frac{486000\,\zeta_3\zeta_5}{49\,\pi^8}\approx1.3029\,.
\end{equation}
Then $p=1$ signifies that the spectrum is thermal up to its second
moment, while $p<1$ signifies a pinched spectrum (high-energy tail
suppressed), and $p>1$ an anti-pinched spectrum (high-energy tail
enhanced). For $p<1.023$ it is common to approximate the spectrum as a
nominal Fermi-Dirac distribution characterized by a temperature $T$
and a degeneracy parameter $\eta$ which are chosen such that
$\langle\epsilon\rangle_{\rm flux}$ and $\langle\epsilon^2\rangle_{\rm
flux}$ are reproduced (Table~\ref{tab:NumericalResultsMesser}). 
Finally we show the neutrino luminosity in the last column of
Table~\ref{tab:NumericalResultsMesser}.

We always include elastic $\nu_\mu {\rm N}$ scattering which provides
the dominant opacity contribution. In a given run we additionally
include those energy-exchanging processes which are indicated in the
first columns of Table~\ref{tab:NumericalResultsMesser}.  We use ``b''
to indicate bremsstrahlung ${\rm NN}\to {\rm NN}\nu_\mu\bar\nu_\mu$,
``r'' for recoil in $\nu_\mu {\rm N}\to {\rm N}\nu_\mu$, ``s'' for
$\nu_\mu{\rm e}^\pm$ scattering, ``p'' for the traditional pair
annihilation process ${\rm e}^+{\rm e}^-\to \nu_\mu\bar\nu_\mu$, ``n''
for the new neutrino annihilation process $\nu_{\rm e}\bar\nu_{\rm
e}\to \nu_\mu\bar\nu_\mu$, and ``sn'' for $\nu_\mu{\rm e}^\pm$ plus
$\numu\nue$ and $\numu\bar\nue$ scattering.

In the first row of Table~\ref{tab:NumericalResultsMesser} we show the
original flux characteristics of our background model from the
calculation of Messer et~al.\ (2002).  Running our code with the same
neutrino reactions---scattering on ${\rm e}^\pm$ (s) and ${\rm e}^+
{\rm e}^-$ pair annihilation (p)---we find the results in the second
row. We attribute the small differences to the different numerical
approaches. In particular, the Boltzmann solver of Messer et~al.\
(2002) works in practice with a limited number of energy bins.
Moreover, there may be differences in the implementation of the
microphysical reactions. Finally, in the Monte Carlo code we assume
that neutrinos are in thermodynamic equilibrium with the stellar
medium at the inner boundary. This choice of boundary condition may
have a small effect on our results since the thermalization depth of
${\rm e}^+ {\rm e}^-$ pair processes is strongly energy dependent and
thus low energy $\numu$ are forced into equilibrium by our boundary
condition instead of the pair process. We interpret the first two rows
of Table~\ref{tab:NumericalResultsMesser} as agreeing satisfactorily
well with each other.

Next we add $\nue \bar\nue$ annihilation (n).  The spectrum remains
almost unchanged, but the luminosity increases by about
30\%. Therefore, the new process has a rather significant impact on
the predicted $\nu_\mu$ luminosity.

However, other processes are also important which in the past have not
been included in numerical simulations. Therefore, we switch off the
new process and instead include bremsstrahlung (b) and recoil (r).
Compared to the original run we obtain almost the same luminosity, but
significantly lowered spectral energies.  Now we again include $\nue
\bar\nue$ annihilation (n) and find that the spectra remain
unaffected, but the luminosity increases by about 20\%.  Therefore,
even with all other energy-exchanging reactions included, the
$\nu_{\rm e} \bar\nu_{\rm e}$ process still has an important effect on
the luminosity.  We finally switch off the traditional pair process
(p), but keep the new one. The spectra remain unaffected, the
luminosity slightly drops. It is evident that the $\nu_{\rm
e}\bar\nu_{\rm e}$ process is by far the dominant leptonic source
reaction for muon neutrinos.  Its importance relative to
bremsstrahlung will depend sensitively on the background model (Keil
et.~al.\ 2002).

As a last step we include the scattering on $\nue$ and $\bar\nue$ in
addition to all other processes. The rate for this process is
typically half as large as the rate for scattering on ${\rm e}^\pm$,
in agreement with Fig.~\ref{fig:easyexpl_scat} if we use $\eta_{\rm e}
\simeq 3$ and $\eta_{\nue} \simeq 0.3$.  The effect on the $\nu_\mu$
flux and spectrum is minimal and in fact below the numerical
resolution of our Monte Carlo runs.

In the second part of Table~\ref{tab:NumericalResultsMesser} we
finally show several runs for the transport of anti-neutrinos. Recall
that $\nu_{\rm e}\bar\nu_{\rm e}\to\nu_\mu\bar\nu_\mu$ generates
different source spectra for $\nu_\mu$ and $\bar\nu_\mu$.  Of course,
the small differences between $\nu_{\mu}$ and $\bar\nu_{\mu}$
scattering off electrons and positrons are also taken into account, as
well as the small differences in ${\rm e}^+ {\rm e}^-$ pair
annihilation.  Comparing the $\bar\nu_\mu$ runs with those for
$\nu_\mu$ we find excellent agreement.  Therefore, the detailed
spectral distribution of the pair rate is not important, only the
total rate of absorption and production of $\nu_\mu\bar\nu_\mu$ pairs
matters.

%%%%%%%%%%%%%%%%%%%%%%%%%%%%%%%%%%%%%%%%%%%%%%%%%%%%%%%%%%%%%%%%%%%%%%
%% Section IV %%%%%%%%%%%%%%%%%%%%%%%%%%%%%%%%%%%%%%%%%%%%%%%%%%%%%%%%
%%%%%%%%%%%%%%%%%%%%%%%%%%%%%%%%%%%%%%%%%%%%%%%%%%%%%%%%%%%%%%%%%%%%%%

\section{\uppercase{Hydrodynamical Simulation 
with Boltzmann Transport}}

\label{sec:Boltzmann}

The large flux increase found by the Monte Carlo method may not
persist in a self-consistent hydrodynamical treatment where the
stellar model can adjust in response to the modified
transport. For this reason we have performed a Boltzmann transport
simulation coupled with a full hydrodynamics code (Rampp
\& Janka 2002) for the 15~$M_{\odot}$ progenitor model s15s7b2
(Woosley, personal communication; Woosley \& Weaver 1995).

In order to minimize the required modifications of the code we treat
the transport of $\nu_\mu$ and $\bar\nu_\mu$ identically.  Therefore,
we use an average source strength for the $\nu_\mu$ and $\bar\nu_\mu$
production from $\nu_{\rm e}\bar\nu_{\rm e}$ annihilation, ignoring
the spectral differences.  The Monte Carlo results suggest that this
approximation is well justified.

For computing the interaction rates we again assume LTE for $\nu_{\rm
e}$ and $\bar\nu_{\rm e}$ in all regions where the new annihilation
process is important. This assumption also justifies introducing an
energy source term for the new process directly into the medium energy
equation in perfect analogy to the source term of the process ${\rm
e}^+{\rm e}^-\lra \nu\bar\nu$.  Similarly, the scattering reactions of
$\nu_{\mu}$ and $\bar\nu_{\mu}$ off $\nu_{\rm e}$ and $\bar\nu_{\rm
e}$ can be treated in full correspondence to the scattering off
electrons and positrons by a simple change of the weak coupling
coefficients. Thus a direct coupling of the $\nu_{\mu}$ and $\nu_{\rm
e}$ sectors of the neutrino transport code is avoided. For the new
process this is achieved only indirectly via the stellar medium as an
intermediary.

\begin{figure}[ht]
\columnwidth=7.5cm
\plotone{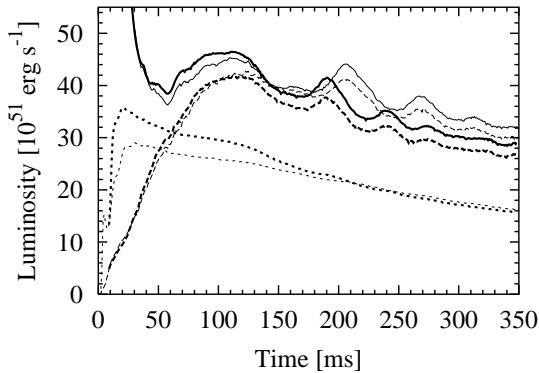}
\caption{\label{fig:lum}Evolution of the luminosities for 
$\nue$ (solid), $\bar\nue$ (dashed) and $\nu_\mu$ (dotted) as
functions of time after bounce for two
hydrodynamical simulations with three-flavor Boltzmann 
neutrino transport. Thick lines represent results of a model
with the new process included, thin lines of a reference 
calculation without this process. The luminosities are
measured by an observer comoving with the stellar fluid
at 500~km.}
\end{figure}

\begin{figure}[ht]
\columnwidth=7.5cm
\plotone{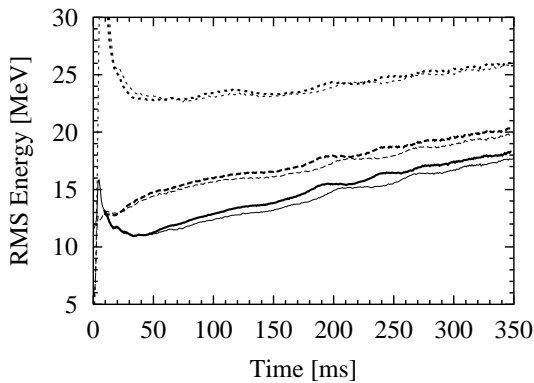}
\caption{\label{fig:erms}Evolution of the rms 
energies for $\nue$ (solid), $\bar\nue$ (dashed) and $\nu_\mu$
(dotted) as measured by an observer comoving with the stellar
fluid at 500~km.}
\end{figure}

\begin{figure}[ht]
\columnwidth=7.5cm
\plotone{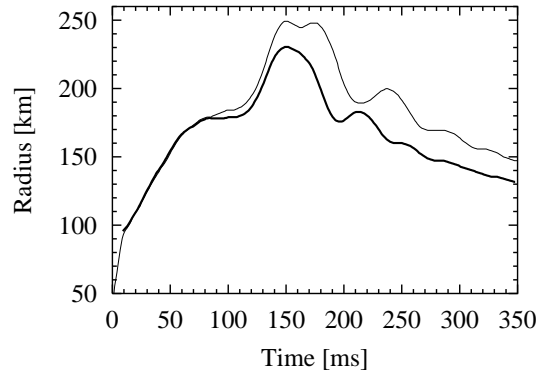}
\caption{\label{fig:spos}Evolution of the shock positions.}
\end{figure}

Our baseline for comparison is a Newtonian simulation which includes
the transport of neutrinos of all three flavors.  Neutrino-medium
interactions for all relevant processes are implemented, notably
nucleon bremsstrahlung, as described by Rampp \& Janka (2002). Note,
however, that nucleon recoils are not taken into account in the
results shown for the baseline simulation (thin lines in
Figs.~\ref{fig:lum}, \ref{fig:erms} and \ref{fig:spos}).  Rather, the
charged-current and neutral-current neutrino reactions with nucleons
are handled in the ``old standard approximation'' of infinitely
massive nucleons at rest following the treatment by Bruenn (1985) and
Mezzacappa \& Bruenn (1993).

In a second simulation we included the new leptonic process for
$\nu_{\mu}\bar\nu_{\mu}$ pair production and annihilation. The results
for this run are depicted as thick lines in the figures.  In
Figs.~\ref{fig:lum} and \ref{fig:erms} we show the evolution of the
neutrino luminosities and of the root mean squared (rms) energies as
functions of time after bounce.
The rms energy is
defined in the
usual way (e.g.\ Messer et~al.\ 1998) by
\begin{equation}
\langle\epsilon\rangle_{\rm rms}\,=\,\sqrt{
{\int_0^\infty\!{\rm d}\epsilon\int_{-1}^{+1}\!{\rm d}\mu\,
f_{\nu}(\epsilon,\mu)\epsilon^5 \over
\int_0^\infty\!{\rm d}\epsilon\int_{-1}^{+1}\!{\rm d}\mu\,
f_{\nu}(\epsilon,\mu)\epsilon^3 } } \,,
\end{equation}
i.e.\ using the neutrino energy distribution as a weight function.
The results are given for an observer who is comoving with the stellar
fluid at a radial location of 500~km. The Doppler-blueshift due to the
infall velocity of the matter is rather small at this distance from
the neutrino emitting neutron star so that the quantities are close to
the observable properties at infinity.

As expected, the $\nu_\mu$ luminosity is increased by 10--20\%, but
the spectrum remains essentially unaffected. The enhanced energy loss
leads to a somewhat faster proto-neutron star contraction.  This
causes small changes also in the region where the electron neutrino
and antineutrino fluxes are built up and where these neutrinos finally
decouple from the stellar background. As a consequence, the mean
spectral energies of $\nu_{\rm e}$ and $\bar\nu_{\rm e}$ are
systematically higher by a small amount (Fig.~\ref{fig:erms}).
Initially, the $\nu_{\rm e}$ luminosity is also slightly larger
(Fig.~\ref{fig:lum}) but after about 150~ms post bounce the
luminosities of $\nu_{\rm e}$ and $\bar\nu_{\rm e}$ drop below the
level of those in the reference simulation. This is caused by the
decrease of the radii of the neutrino spheres in response to the
accelerated contraction of the proto-neutron star.  At 200~ms after
bounce the $\numutau$ luminosities of the two models then become
nearly equal, probably as a consequence of two competing effects which
seem to essentially compensate each other at later times: On the one
hand side the neutrino emission of the nascent neutron star decays
faster with time when the new process is included (as visible from the
$\nue$ and $\bar\nue$ luminosities), on the other hand the new
process raises the energy loss in $\numutau$ compared to $\nue$ and
$\bar\nue$.

The shock positions in both simulations evolve identically until about
100~ms after bounce. Then the shock is somewhat weakened by
introducing the new reaction and expands only to a maximum radius of
230~km instead of 250~km (Fig.~\ref{fig:spos}). This can be understood
again by the more compact proto-neutron star, which causes higher
infall velocities in the region of heating by $\nu_{\rm e}$ and
$\bar\nu_{\rm e}$ absorption behind the shock. This reduces the
integral energy deposition by neutrinos as well as the pressure behind
the shock front.


Since we completed this study we have implemented the neutrino
annihilation process in simulations with an approximate treatment of
general relativity (Rampp \& Janka 2002). In these new simulations
neutrino-nucleon interactions are implemented in terms of dynamical
structure functions for correlated nuclear matter and also include
weak magnetism corrections (Rampp et al.\ 2002). Put another way,
these simulations include nuclear recoil as well as nuclear
correlation effects.  The impact of the new process in these more
complete simulations is comparable to what we have found here.

%%%%%%%%%%%%%%%%%%%%%%%%%%%%%%%%%%%%%%%%%%%%%%%%%%%%%%%%%%%%%%%%%%%%%%
%% Section V %%%%%%%%%%%%%%%%%%%%%%%%%%%%%%%%%%%%%%%%%%%%%%%%%%%%%%%%%
%%%%%%%%%%%%%%%%%%%%%%%%%%%%%%%%%%%%%%%%%%%%%%%%%%%%%%%%%%%%%%%%%%%%%%

\section{\uppercase{Conclusions}}

\label{sec:Conclusions}

We have shown that $\nue \bar\nue$ annihilation is the dominant
leptonic source for $\nu_{\mu,\tau}\bar\nu_{\mu,\tau}$ pairs in a SN
core, far more important than the traditional ${\rm e}^+ {\rm e}^-$
pair annihilation. Its importance relative to the nucleon
bremsstrahlung process, which also has not been included in previous
SN simulations, depends on the stellar profile. In our Monte Carlo
studies with a background model representing the SN core during the
accretion phase, the new process enhanced the $\nu_\mu$ luminosity by
about 20\% in a calculation where nucleon bremsstrahlung and energy
exchange by nucleon recoils were taken into account. Without these
other new effects, $\nue \bar\nue$ annihilation has an even larger
impact.

In self-consistent hydrodynamical simulations with Boltzmann neutrino
transport we find that during the first 150~ms the effect is similar
to that obtained in the static Monte Carlo simulations and has a
noticeable influence on the stellar evolution and structure.  Later,
the $\nu_\mu$ and $\bar\nu_\mu$ luminosities approach those of the
model without the new process of $\nue \bar\nue$ annihilation.  At
that time, however, the $\nu_{\rm e}$ and $\bar\nu_{\rm e}$
luminosities are smaller.  Throughout the simulation the new reaction
therefore works in the direction of making the $\nu_{\rm e}$ and
$\bar\nu_{\rm e}$ luminosities more similar to those of $\nu_\mu$ and
$\bar\nu_\mu$.  In the model with the new process the shock is
somewhat weakened and reaches only smaller radii.

In all of our Monte Carlo and hydrodynamical runs we confirm the naive
expectation that the effect of the new pair production and
annihilation process on the neutrino spectra is minimal.  The crossed
process of $\nu_{\mu,\tau}$ scattering off $\nu_{\rm e}$ and
$\bar\nu_{\rm e}$ also turned out to have a negligible impact on the
emitted spectra.

We conclude that state-of-the-art SN simulations should include the
$\nue \bar\nue$ annihilation reaction to
$\nu_{\mu,\tau}\bar\nu_{\mu,\tau}$ pairs.  While the effects of this
process on the neutrino luminosities and spectra and on the shock
propagation are not dramatic, they are nevertheless noticeable and not
negligible, even after nucleon bremsstrahlung and nucleon recoils have
been included.  Implementing the new process is not more difficult and
not more CPU-expensive than the traditional ${\rm e}^+ {\rm e}^-$
process.

%%%%%%%%%%%%%%%%%%%%%%%%%%%%%%%%%%%%%%%%%%%%%%%%%%%%%%%%%%%%%%%%%%%%%%
%% Acknowledgments %%%%%%%%%%%%%%%%%%%%%%%%%%%%%%%%%%%%%%%%%%%%%%%%%%%
%%%%%%%%%%%%%%%%%%%%%%%%%%%%%%%%%%%%%%%%%%%%%%%%%%%%%%%%%%%%%%%%%%%%%%

\section*{\uppercase{Acknowledgements}}

\label{sec:Acknowledgements}

We thank Bronson Messer for making the unpublished results of a
collapse simulation available.  This work was partly funded by the
Deutsche Forschungsgemeinschaft under grant No. SFB 375 and the ESF
network Neutrino Astrophysics. The radiation-hydrodynamical
computations were performed on the NEC SX-5/3C of the Rechenzentrum
Garching and the CRAY T90 of the John von Neumann Institute for
Computing (NIC) in J\"ulich, Germany.

%%%%%%%%%%%%%%%%%%%%%%%%%%%%%%%%%%%%%%%%%%%%%%%%%%%%%%%%%%%%%%%%%%%%%%
%% References %%%%%%%%%%%%%%%%%%%%%%%%%%%%%%%%%%%%%%%%%%%%%%%%%%%%%%%%
%%%%%%%%%%%%%%%%%%%%%%%%%%%%%%%%%%%%%%%%%%%%%%%%%%%%%%%%%%%%%%%%%%%%%%

\end{document}